\documentclass[12pt]{article}
\usepackage{epsfig,amsmath,amssymb,mathrsfs}
\usepackage[utf8]{inputenc}
\usepackage{feynmp-auto}
\usepackage[colorlinks=true,citecolor=blue,,linktocpage=true,linkcolor=blue,urlcolor=black]{hyperref}
  {\end{list}}%

\makeatletter
\renewcommand{\section}{\setcounter{equation}{0}\@startsection
 {section}%
 {1}%
 {0pt}%
 {-1\baselineskip}%
 {0.4\baselineskip}%
 {\bfseries\large}}%
\renewcommand{\subsection}{\@startsection
 {subsection}%
 {2}%
 {0pt}%
 {-0.75\baselineskip}%
 {0.2\baselineskip}%
 {\bfseries}}%
\renewcommand{\subsubsection}{\@startsection
 {subsubsection}%
 {3}%
 {0pt}%
 {-0.5\baselineskip}%
 {0.1\baselineskip}%
 {\sc}}%
\makeatother

\DeclareMathAlphabet{\mathpzc}{OT1}{pzc}{m}{it}

\def\a{\alpha}

\def\g5{\gamma_{5}}

\def\m{\mu}
\def\n{\nu}
\def\r{\rho}
\def\s{\sigma}

\def\Aslash{{A\mkern-11mu/}}

\def\kslash{{k\mkern-8mu/}{\!}}
\def\prslash{{\partial\mkern-9mu/}}

\def\prslash{{\partial\mkern-9mu/}}    
\def\qslash{{q\mkern-8mu/}{\!}}

\def\idkn{\int\!\! \frac{d^{n}k\!}{(2\pi)^{n}}}

\def\idx{\int\!\! d^4\!x}

\def\idxn{\int\!\! d^{n}\!x}
\def\idyn{\int\!\! d^{n}\!y}

\def\idxonen{\int\!\! d^{n}\!x_1}
\def\idxtwon{\int\!\! d^{n}\!x_2}
\def\idxthreen{\int\!\! d^{n}\!x_3}
\def\idx{\int\!\! d^{4}\!x}

\newcommand{\bea}{\begin{eqnarray}}
\newcommand{\eea}{\end{eqnarray}}
\newcommand{\beann}{\begin{eqnarray*}}
\newcommand{\eeann}{\end{eqnarray*}}
\newcommand{\ba}{\begin{array}}
\newcommand{\ea}{\end{array}}

 \def\g {\gamma}

\begin{document}
 \begin{titlepage}
\rightline{\scriptsize{FTI/UCM 37-2017}}
\vglue 50pt

\begin{center}

{\Large \bf Unimodular Gravity and  the lepton anomalous magnetic moment at one-loop}\\
\vskip 1.0true cm {Carmelo P. Mart\'{\i}n\footnote{E-mail:
carmelop@fis.ucm.es}
\vskip 0.3 true cm
\it Departamento de F\'{\i}sica Te\'orica I,
Facultad de Ciencias F\'{\i}sicas\\
Universidad Complutense de Madrid,
 28040 Madrid, Spain}\\

\vskip 1.2 true cm

\vskip 0.85 true cm
{\leftskip=50pt \rightskip=50pt \noindent
We work out the one-loop contribution to the lepton anomalous magnetic moment coming from Unimodular Gravity. We use Dimensional Regularization and Dimensional Reduction to carry out the computations. In either case, we find that Unimodular Gravity gives rise  to the same one-loop correction as that of General Relativity.
\par}
\end{center}

\vspace{9pt}
\vfill
\end{titlepage}

\section{Introduction}

It has long been known ~\cite{vanderBij:1981ym} that the most economical way~\cite{Alvarez:2006uu, Alvarez:2005iy, Barcelo:2014mua} --ie, by introducing a minimum number of infinitesimal independent gauge redundancies-- to formulate a quantum field theory of spin-two massless particles --ie, gravitons-- is to take the group of transverse  diffeomorphisms as the gauge group of the field theory --rather than the full diffeomorphism group-- and, postulate that the configuration space of the theory is the set of Lorentzian unimodular --ie, $-\rm{det}( g_{\mu\nu})=1$-- metrics. The theory of gravity so obtained is called Unimodular Gravity.

If one thinks --drawing upon lessons learned by using the modern on-shell techniques~\cite{Arkani-hamed:2016gr}--  of gauge symmetries as unphysical redundancies introduced within the standard framework --say, the path integral-- of quantum field theory to make manifest locality and unitarity, it would be conceptually  advisable to formulate, within the standard framework, quantum field theories by using the minimum possible number of such redundancies. Unimodular Gravity is a concrete realization of this approach to formulating a quantum field theory of gravitons. In this case,
the case of Unimodular Gravity, to follow the approach in question has physical consequences: the energy of the vacuum does not gravitate --thus, a breach of the equivalence principle occurs-- and, therefore,  the question of why
the Cosmological Constant is not of the order of $M_{pl}^2$, $M_{pl}$ being the Planck mass, does not arise~\cite{Zee:2013dea, Unruh:1988in, Henneaux:1989zc, Wilczek:1998ea, Smolin:2009ti}. This way,
ie, by removing the problem, one of the problems involving the Cosmological Constant is solved; there still remains the issue of why it has its astronomically observed value.

Several aspects of Unimodular Gravity have been studied recently. These range from the study of classical black-hole solutions~\cite{Oda:2016knt, Oda:2016nvc, Chaturvedi:2016fea} to its Cosmological implications~\cite{Gao:2014nia, Cho:2014taa} --incluiding its consistency with the Inflation paradigm~\cite{Ellis:2013eqs}, along with the analysis of the UV structure of it~\cite{Saltas:2014cta,Eichhorn:2015bna, Alvarez:2015sba}  and some formal issues~\cite{Kluson:2014esa,Bufalo:2015wda,Alvarez:2015oda, Alvarez:2016lbz}.

Classically there is no difference~\cite{Ellis:2010uc} between General Relativity and Unimodular Gravity, provided we use in Unimodular Gravity an energy-momentum tensor which is covariantly conserved --see~\cite{Josset:2016vrq}, for a
proposal with a nonconserved energy-momentum tensor. The question arises as to whether Unimodular Gravity differs from General Relativity at the quantum level, when they are understood as effective quantum field theories. To  answer this question with full generality is  a highly non trivial issue since both the gauge group  and the configuration space of fields  change when one moves from one theory to the other  and, further, both the diffeomorphisms and gravitational field --the metric- to be used are constrained --leading to the introduction of ghosts of ghosts-- in the case of Unimodular Gravity --see~\cite{Alvarez:2015sba}, for further details. Hence, it is worth redoing, using Unimodular Gravity instead, the classic one-loop computations done long ago within the framework quantum General Relativity formulated as a quantum  field theory and look for differences, if any. In~\cite{Alvarez:2015sba}, we carried out the counterpart computation of the classic computation done in~\cite{tHooft:1974toh} within quantum General Relativity. The conclusion was that, as in the General Relativity case, there are no  on-shell
UV divergent corrections to the one-loop effective action. Off-shell there are UV divergent contributions, which do not match those of General Relativity.
This is not a surprise since those off-shell contributions are gauge dependent. Agreement between General Relativity and Unimodular Gravity is
only expected, if there is any, when physical quantities are compared. This we did in~\cite{Alvarez:2016uog} by computing some tree-level MHV amplitudes and found complete agreement between the Unimodular Gravity and the General Relativity results.

Another classic computation is the one of the one-loop quantum General Relativity contribution to the anomalous magnetic moment of a lepton, which was carried out in~\cite{Berends:1974tr}. It turned out that although each Feynman diagram contribution was UV divergent --indeed, each Feynman diagram is not UV finite by powercounting-- the net result is UV finite, a phenomenon explained in~\cite{Bellucci:1984dx}. However, in the absence of Supergravity,  the cancellation of the UV divergent piece that occurs in the sum over all Feynman diagram contributions leaves a net finite remnant. This finite remnant is regularization dependent --as shown in \cite{Wilcox:1981ff, delAguila:1984fv, Grisaru:1984ps,delAguila:1997yd}, and, hence, it cannot be taken  as an observable correction to $(g-2)$-factor coming from the Standard Model.
Actually, in the absence of Supersymmetry, the following finite counterterm
\begin{equation*}
C \frac{m^2}{M_{pl}^2}\,\idx\; \bar{\psi}(x)\,\frac{i\sigma^{\mu\nu}}{2m}\,\psi(x)\,F_{\m\nu}(x),
\end{equation*}
where $M_{pl}$ is the Planck Mass, must be added to tree-level action, if General Relativity is defined, at the quantum level, as an effective field theory~\cite{Donoghue:2017pgk}. This tree-level counterterm yields a contribution to $(g-2)$-factor of the lepton that must be added to the one-loop finite correction computed by   using a given regularization  method --as done in~\cite{Berends:1974tr, Wilcox:1981ff, delAguila:1984fv, Grisaru:1984ps, delAguila:1997yd}-- and, then, fixing the $C$ parameter by matching with an UV complete  theory or  by taking $C$ from experiment. In either case, the contribution to the lepton $(g-2)$-factor
coming from one-loop loop diagrams involving the exchange of one graviton has to be computed by using a given regularization method, as was done, for instance, in ~\cite{Berends:1974tr}. The purpose of this paper is to carry out such computation when, instead of General Relativity, one uses Unimodular Gravity to define an effective quantum field theory of Gravity. We shall carry out the computation by using the BRS quantization procedure laid out in~\cite{Alvarez:2015sba} and regularize the resulting Feynman diagrams by employing both Dimensional Regularization and Dimensional Reduction --the two most
widely used regularization methods.

Let us stress that since the finite contributions found in~\cite{Berends:1974tr} --for Dimensional Regularization-- and~\cite{delAguila:1984fv} --for Dimensional Reduction-- are not observables of the theory, as we have discussed above, there is no clear reason why one should obtain the corresponding same results when Unimodular Gravity is used to set a quantum field theory of Gravity, instead of General Relativity.

The lay out of this paper is as follows. In Section 2, we display the QED action coupled to Unimodular Gravity and expand it around the Minkowski metric up first order in the graviton field. Section 3 is devoted to the computation of the one-loop Unimodular Gravity contributions lepton $(g-2)$-factor. In this section we show the there is no new contributions as compared to those of General Relativity. Section 4 is devoted to the analysis of the null effect that the contributions produced by the contact terms in the Ward identity governing the energy-momentum conservation in our  quantum field theory has on the lepton $(g-2)$-factor. In the Appendix we discuss how we have carried out the computation of the net contributions coming from some tricky IR divergent Feynman integrals.

\section{QED coupled to Unimodular Gravity: the action}

Let $\hat{g}_{\mu\nu}$ denote the Unimodular --ie, with determinant equal to -1-- metric of the $n$ dimensional spacetime manifold. We shall assume the mostly minus signature for the metric.

Let $\eta_{ab}$ denote the Minkowski metric on the tangent vector space at each point of the spacetime manifold. Let $\hat{e}^a_{\mu}$ denote the vielbein for the unimodular metric $\hat{g}_{\mu\nu}$ and let $\hat{e}^{\mu}_a$ be the inverse of $\hat{e}^a_{\mu}$:
\begin{equation}
\begin{array}{l}
{\hat{e}^a_{\mu}\hat{e}^b_{\nu}\eta_{ab}=\hat{g}_{\mu\nu},\quad \hat{e}^a_{\mu}\hat{e}^b_{\nu}\hat{g}^{\mu \nu}=\eta^{ab},}\\[8pt]
{\hat{e}^a_{\mu}\hat{e}^{\mu}_b=\delta^a_b,\quad \hat{e}^a_{\nu}\hat{e}^{\mu}_a=\delta^{\mu}_{\nu}.}
\label{vielbein}
\end{array}
\end{equation}
The torsion-free spin connection,$\hat{\omega}_{\mu}$,  for the metric $\hat{g}_{\mu\nu}$ is given by the formula:
\begin{equation}
\hat{\omega}_{\mu}=\frac{1}{8}[\gamma^a,\gamma^b]\eta_{ac}\hat{e}^c_{\nu}\hat{\nabla}_{\mu}\hat{e}^{\nu}_b,
\label{spincon}
\end{equation}
$\hat{\nabla}_{\mu}$ stands for the covariant derivative for the metric $\hat{g}_{\mu\nu}$ and the $\gamma^a$'s are the Dirac matrices:
\begin{equation*}
\{\gamma^a,\gamma^b\}=2\eta^{ab}.
\end{equation*}

The classical action,$S$, of QED coupled to Unimodular Gravity in $n$ dimensions reads
\begin{equation}
\begin{array}{l}
{S\;=\;S_{UG}\,+\,S_{QED}}\\[8pt]
{S_{UG}=-\frac{2}{\kappa^2}\idxn\; R[\hat{g}_{\mu\nu}]}\\[8pt]
{S_{QED}\,=\,-\frac{1}{4}\idxn\,\hat{g}_{\mu\rho}\hat{g}_{\nu\lambda}\,F^{\mu\nu}F^{\rho\lambda}+\idxn\;\bar{\psi}(i\hat{e}^{\mu}_{a}\gamma^a D[A]_{\mu}-m)\psi},
\end{array}
\label{classaction}
\end{equation}
where $\kappa^2=32\pi G$, $R[\hat{g}_{\mu\nu}]$ is the scalar curvature for the unimodular metric,
$F_{\mu\nu}=\partial_\mu A_\nu-\partial_\nu A_\mu$ and
\begin{equation*}
D[A]_{\mu}=\partial_{\mu}+\hat{\omega}_{\mu}+i\,e\,A_{\mu}.
\end{equation*}

To quantize the theory we shall proceed as in Refs.~\cite{Alvarez:2005iy,Alvarez:2006uu, Alvarez:2015sba} and introduce the unconstrained fictitious metric, $g_{\mu\n}$, thus
\begin{equation}
\hat{g}_{\mu\nu}=(-g)^{-1/n}\,g_{\mu\nu};
\label{fictitious}
\end{equation}
where $g$ is the determinant of $g_{\mu\nu}$. Then, we shall express the action in (\ref{classaction}) in terms of the fictitious metric $g_{\mu\nu}$ by using  (\ref{fictitious}) and, finally, we shall define the path integral by integration over $g_{\mu\nu}$ and the matter fields; once an appropriate BRS invariant action  has been constructed.

To compute the Unimodular Gravity one-loop contributions to the lepton $(g-2)$-factor, one splits $g_{\mu\nu}$ as follows
\begin{equation}
g_{\mu\nu}=\eta_{\mu\nu}\,+\,\kappa h_{\mu\nu}
\label{gsplitting}
\end{equation}
and, then, expands $S$, in (\ref{classaction}), up to terms quadratic in $h_{\mu\nu}$. Actually, the terms coming from $S_{QED}$ which are quadratic in $h_{\mu\nu}$ are not needed in Dimensional  Regularization (or Dimensional Reduction), since they give rise to one-loop contributions to the lepton $(g-2)$-factor which only involve integrals over the loop-momentum of the type
\begin{equation*}
\idkn\,\frac{k_{\mu}...k_{\rho}}{(k^2)^m} ,
\end{equation*}
and all these integrals vanish.

Taking into account (\ref{vielbein}), (\ref{spincon}), (\ref{fictitious}) and (\ref{gsplitting}), one obtains
\begin{equation*}
\begin{array}{l}
{\hat{e}^a_{\mu}=\delta^a_{\mu}+\kappa C^a_{\mu}+ O(\kappa^2),\quad C^a_\mu=\frac{1}{2}(h_{\mu\nu}-\frac{1}{n}h_\rho^\rho\eta_{\mu\nu})\eta^{\nu a},}\\[8pt]
{\hat{\omega}_\mu=\frac{\kappa}{4}[\gamma^a,\gamma_b]\delta_a^\nu\partial_\nu C^b_\mu\,+\,O(\kappa^2)};
\end{array}
\end{equation*}
and, hence,
\begin{equation}
S_{QED}=-\frac{1}{4}\idxn\,F_{\mu\nu}F^{\mu\nu}+\idxn\;\bar{\psi}(i\prslash-e\Aslash-m)\psi
-\frac{\kappa}{2}\idxn\,T^{\mu\nu}\hat{h}_{\mu\nu}+O(\kappa^2),
\label{sqedexpanded}
\end{equation}
where all the indices in the previous expression are flat indices --ie, with regard to the Minkowski metric, $\eta_{\mu\nu}$-- and
\begin{equation}
\begin{array}{l}
{\hat{h}_{\mu\nu}=h_{\mu\nu}-\frac{1}{n}h^\rho_\rho\eta_{\mu\nu}}\\[8pt]
{T^{\mu\nu}=\frac{i}{4}\bar{\psi}(\gamma^\mu\partial^\nu+\gamma^\nu\partial^\mu)\psi-\frac{i}{4}\bar{\psi}(\gamma^\mu\partial^\nu+\gamma^\nu\partial^\mu)\psi
+F^{\mu}_{\phantom{\mu}\rho} F^{\rho\nu}+\frac{1}{4}F_{\sigma\rho}F^{\sigma\rho}\eta^{\mu\nu}-\frac{e}{2}\bar{\psi}(\gamma^\mu A^\nu+\gamma^\nu A^\mu)\psi}.
\end{array}
\label{Tandhathtensors}
\end{equation}
Notice that $T^{\mu\nu}$ above is the energy-momentum tensor of Ref.~\cite{Berends:1974tr}, but, now, unlike General Relativity, it does not couple to the full graviton field, $h_{\mu\nu}$, but only to its traceless part, $\hat{h}_{\mu\nu}$.

In addition to the Feynman rules  derived from $S_{QED}$ in (\ref{sqedexpanded}), we shall also need the free graviton propagator, which, for the gauge choice of Ref.~\cite{Alvarez:2015sba}, reads
\begin{equation}
\begin{array}{l}
{\langle h_{\mu\nu}(k)h_{\rho\sigma}(-k)\rangle=}\\[8pt]
{\frac{i}{2k^2}\left(\eta_{\m\s}\eta_{\n\r}+\eta_{\m\r}\eta_{\n\s}\right)-\frac{i}{k^2}\frac{\a^2n^2-n+2}{\a^2 n^2(n-2)}\eta_{\m\n}\eta_{\r\s}+\frac{2i}{n-2}\left(\frac{k_\r k_\s \eta_{\m\n}}{(k^2)^2}+\frac{k_\m k_\n \eta_{\r\s}}{(k^2)^2}\right)-\frac{2in}{n-2}\frac{k_{\m}k_{\n}k_{\r}k_{\s}}{(k^2)^3}
}
\end{array}
\label{propagatorug}
\end{equation}	
Actually, it is only the propagator of the traceless part, $\hat{h}_{\mu\nu}$, of the graviton field, the object which enters our computations. This propagator, which is easily derived by using (\ref{propagatorug}), runs thus
\begin{equation}
\hat{\Delta}_{\mu\nu,\rho\sigma}(k)=\langle\hat{h}_{\mu\nu}(k)\hat{h}_{\rho\sigma}(-k)\rangle=\Delta^{(GR)}_{\mu\nu,\rho\sigma}(k)+\Delta_{\mu\nu,\rho\sigma}(k),
\label{tracelessprop}
\end{equation}
where
\begin{equation}
\begin{array}{l}
{\Delta^{(GR)}_{\mu\nu,\rho\sigma}(k)=\frac{i}{2k^2}\left(\eta_{\m\s}\eta_{\n\r}+\eta_{\m\r}\eta_{\n\s}-\frac{2}{n-2}\eta_{\m\n}\eta_{\r\s}\right)
,}\\[8pt]
{\Delta_{\mu\nu,\rho\sigma}(k)=\frac{2i}{n-2}\frac{k_{\mu}k_{\nu}\eta_{\rho\sigma}+k_{\rho}k_{\sigma}\eta_{\mu\nu}}{(k^2)^2}-
\frac{2in}{n-2}\frac{k_{\mu}k_{\nu}k_{\rho}k_{\sigma}}{(k^2)^3}.
}
\end{array}
\label{Deltasdef}
\end{equation}	
Notice that $\Delta^{(GR)}_{\mu\nu,\rho\sigma}(k)$ is the free graviton propagator of General Relativity in the de Donder gauge. The de Donder gauge is the gauge used in ~\cite{Berends:1974tr} to carry out the computations.

\section{Computing the Unimodular Gravity contributions to the lepton $(g-2)$-factor}

Let us consider the following matrix element
\begin{equation}
\begin{array}{l}
{\frac{\kappa^2}{8}\idxonen\idxtwon\idxthreen\idxn\idyn\,e^{-i(p x_1 + q x_3-p' x_2)}}\\[8pt]
{\bar{u}(p')(i\overrightarrow{\prslash_{x_2}}-m)\langle \psi(x_2)\bar{\psi}(x_1) A^{\lambda}(x_3)
T^{\mu\nu}(x)T^{\rho\sigma}(y)\rangle(-i\overleftarrow{\prslash_{x_1}}-m)u(p)\hat{\Delta}_{\mu\nu,\rho\sigma}(x-y),}
\end{array}
\label{matrixelement}
\end{equation}
where $p^2=m^2$ and $(p')^2=m^2$ and $q^2<0$, for QED in Minkowski spacetime and in the Feynman gauge. $T^{\mu\nu}(x)$ is given in (\ref{Tandhathtensors})
and $\Delta_{\mu\nu,\rho\sigma}(x-y)$ is the free propagator of  $\hat{h}_{\mu\nu}$, whose Fourier transform is given in (\ref{propagatorug}).

Then, the one-loop Unimodular Gravity contributions to the lepton $(g-2)$-factor, $(g-2)_l$, are obtained by extracting first from the matrix element in (\ref{matrixelement}) the contributions that are of the form
\begin{equation}
-i\,e\,m^2\kappa^2\,F_2\Big(\frac{q^2}{m^2}\Big)\, \bar{u}(p')\,\frac{i\sigma^{\lambda\rho}q_\rho}{2m}\,u(p),\quad \sigma^{\lambda\rho}=\frac{i}{2}[\gamma^\lambda,\gamma^\rho],
\label{anomalousterms}
\end{equation}
and then taking the limit $q^2/m^2\rightarrow 0$ of $F_2\big(q^2/m^2\big)$ as $q^2<0$:
\begin{equation}
(g-2)^{UG}_l\,=\,2\, m^2\kappa^2\,\lim_{q^2/m^2\rightarrow 0^{-}}\,F_2\Big(\frac{q^2}{m^2}\Big).
\label{g2lfactor}
\end{equation}

The Feynman diagrams which represent the contributions to the matrix element in (\ref{matrixelement}) which give rise to the Unimodular Gravity contribution to the lepton $(g-2)$-factor are depicted in Figures 1, 2 and 3. In those diagrams the continuous lines represent  leptons, the single wavy lines photons and the double wavy lines the free propagator, $\hat{\Delta}_{\mu\nu,\rho\sigma}(k)$, of the traceless field $\hat{h}_{\mu\nu}$. It is understood that each fermion external leg of those diagrams is to be replaced with  $u(p)$, if it is ingoing, or with $\bar{u}(p')$, if it is outgoing. The photon external leg is amputated. Notice that the only difference between the diagrams in Figures 1, 2 and 3 and those displayed in Ref.~\cite{Berends:1974tr} is the value of the graviton line, for the $T^{\mu\nu}(x)$ entering the matrix element in (\ref{matrixelement}) is the same energy-momentum tensor employed in Ref.~\cite{Berends:1974tr}.

Now, recall that $\hat{\Delta}_{\mu\nu,\rho\sigma}(k)$ is the sum of the free graviton propagator of General Relativity in the de Donder gauge and $\Delta_{\mu\nu,\rho\sigma}(k)$, as displayed in (\ref{tracelessprop}) and (\ref{Deltasdef}). Hence, it is plain that if there is any different between the General Relativity contribution to the lepton $(g-2)$-factor and the Unimodular Gravity contribution to the latter, it can be obtained by replacing $\hat{\Delta}_{\mu\nu,\rho\sigma}(k)$ with $\Delta_{\mu\nu,\rho\sigma}(k)$ in the matrix element
in (\ref{matrixelement}). So, to compute such difference, if any, it is enough to assume that the double wavy line in the diagrams in Figures 1, 2 and 3, represents
$\Delta_{\mu\nu,\rho\sigma}(k)$, instead of $\hat{\Delta}_{\mu\nu,\rho\sigma}(k)$, and, then, extract from them the contributions of the type displayed in (\ref{anomalousterms}) and
(\ref{g2lfactor}). This we have done --in the Appendix we display some detailed computations-- and obtained the following results:

\noindent
\underline{Contribution from the diagram in Figure 1:}
\begin{equation}
-ie \frac{m^2\kappa^2}{8}\frac{1}{n-2}\frac{1}{16 \pi^2}\big[2(\frac{1}{\epsilon}-\gamma-\ln(\frac{m^2}{4\pi\mu^2})+3)\big]\bar{u}(p')\,\frac{i\sigma^{\lambda\rho}q_\rho}{2m}\,u(p)
\label{anomalous1}
\end{equation}
\noindent
\underline{Contribution from the diagrams in Figure 2:}
\begin{equation}
-ie \frac{m^2\kappa^2}{8}\frac{2}{n-2}\frac{1}{16 \pi^2}\big[2(\frac{1}{\epsilon}-\gamma-\ln(\frac{m^2}{4\pi\mu^2})+3)\big]\bar{u}(p')\,\frac{i\sigma^{\lambda\rho}q_\rho}{2m}\,u(p)
\label{anomalous2}
\end{equation}

\noindent
\underline{Contribution from the diagrams in Figure 3:}
\begin{equation}
-ie \frac{m^2\kappa^2}{8}\frac{2}{n-2}\frac{1}{16 \pi^2}\big[-4(\frac{1}{\epsilon}-\gamma-\ln(\frac{m^2}{4\pi\mu^2})+3)\big]\bar{u}(p')\,\frac{i\sigma^{\lambda\rho}q_\rho}{2m}\,u(p)
\label{anomalous3}
\end{equation}
In the previous expressions $\epsilon$, defined by $n=4-2\epsilon$, is regulator and $\mu$ is the renormalization scale. The results in (\ref{anomalous1}), (\ref{anomalous2}) and
(\ref{anomalous3}) have been obtained both with Dimensional Regularization and Dimensional Reduction. This is unlike the contributions coming from General Relativity --see\cite{delAguila:1984fv}-- whose values in Dimensional Regularization and Dimensional Reduction differ. Let us point out that the term in $\Delta_{\mu\nu,\rho\sigma}(k)$ which contains four momenta yields a vanishing contribution to the lepton $(g-2)$-factor for each diagram separately.

Notice that each individual contribution in (\ref{anomalous1}), (\ref{anomalous2}) and
(\ref{anomalous3}) is UV divergent and contains a finite part. And yet, when one adds them all they yield a vanishing contribution. We thus conclude that Unimodular Gravity gives the
same one-loop contribution to lepton $(g-2)$-factor as General Relativity does. Let us stress that the non-vanishing contributions in (\ref{anomalous1}), (\ref{anomalous2}) and
(\ref{anomalous3}) tell us that the equivalence between Unimodular Gravity and General Relativity does not generally occur diagram to diagram but after summing over them.

Before we close this section, we would like to warn the reader that the simplicity of the results in (\ref{anomalous1}), (\ref{anomalous2}) and (\ref{anomalous3}) may convey the false impression that the computations leading  to them are technically trivial. This is not so, for many of the Feynman integrals arising in their computation are IR divergent --the full one-loop vertex itself not being IR finite-- when the momentum, $q$, of the external photon is set to zero, and, then, the computation of the limit in (\ref{g2lfactor}) demands the use of appropriate methods of mathematical analysis --see the Appendix.

\vspace{2cm}

\newcommand{\marrow}[5]{%
    \fmfcmd{style_def marrow#1
    expr p = drawarrow subpath (1/5, 3/5) of p shifted 6 #2 withpen pencircle scaled 0.4;
    label.#3(btex #4 etex, point 0.4 of p shifted 6 #2);
    enddef;}
    \fmf{marrow#1,tension=0}{#5}}
\newcommand{\marrowdou}[5]{%
    \fmfcmd{style_def marrowdou#1
    expr p = drawarrow subpath (1/5, 3/5) of p shifted 6 #2 withpen pencircle scaled 0.1;
    label.#3(btex #4 etex, point 0.1 of p shifted 6 #2);
    enddef;}
    \fmf{marrowdou#1,tension=0}{#5}}
\unitlength = 3mm
\begin{figure}[h!]
\centering
\begin{fmffile}{onephotontriangleDiagram}
    \begin{fmfgraph*}(20,20)
        \fmfleft{i1}
        \fmftop{j1}
        \fmfright{o1}
        \fmf{fermion}{i1,v1,v2,v3,o1}
        \fmf{dbl_wiggly}{v1,v3}
        \fmf{photon}{v2,j1}
\marrow{a}{down}{bot}{$p$}{i1,v1}
\marrow{c}{left}{lft}{$q$}{j1,v2}
\marrow{d}{down}{bot}{$p'$}{v3,o1}
\fmflabel{$e^-$}{i1}
        \fmflabel{$e^-$}{o1}
        \fmflabel{$\lambda$}{j1}
    \end{fmfgraph*}
    \end{fmffile}
    \vspace{-3cm}
 \caption{Triangle Diagram}
    \end{figure}
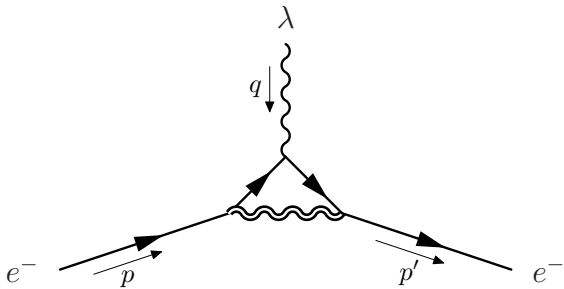

\begin{figure}[h!]
\centering
\begin{fmffile}{blobdiagram}
    \begin{fmfgraph*}(20,20)
        \fmfleft{i1}
        \fmftop{j1}
        \fmfright{o1}
        \fmf{dbl_wiggly,left,tension=0.5}{v1,v2}
        \fmf{fermion}{i1,v1,v2,o1}
        \fmf{photon}{v1,j1}
    \marrow{a}{down}{bot}{$p$}{i1,v1}
\marrow{c}{left}{lft}{$q$}{j1,v1}
\marrow{d}{down}{bot}{$p'$}{v2,o1}
      \fmflabel{$e^-$}{i1}
        \fmflabel{$e^-$}{o1}
        \fmflabel{$\lambda$}{j1}
    \end{fmfgraph*}
  \hskip 2cm
 \begin{fmfgraph*}(20,20)
        \fmfleft{i1}
        \fmftop{j1}
        \fmfright{o1}
        \fmf{fermion}{i1,v1,v2,o1}
        \fmf{dbl_wiggly,left,tension=0.5}{v1,v2}
        \fmf{photon}{v2,j1}
    \marrow{a}{down}{bot}{$p$}{i1,v1}
\marrow{c}{left}{lft}{$q$}{j1,v2}
\marrow{d}{down}{bot}{$p'$}{v2,o1}
      \fmflabel{$e^-$}{i1}
        \fmflabel{$e^-$}{o1}
        \fmflabel{$\lambda$}{j1}
    \end{fmfgraph*}
\end{fmffile}
 \vspace{-3cm}
\caption{Blob Diagrams}
\end{figure}
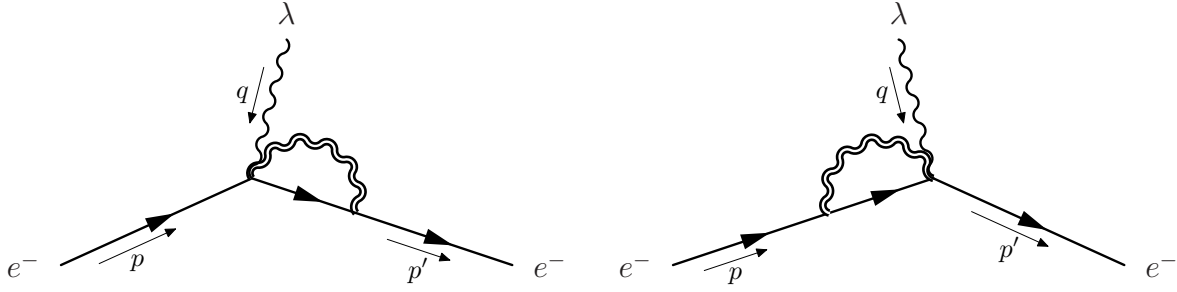

\begin{figure}[h!]
	    	 	\centering
\begin{fmffile}{twophoton}
    \begin{fmfgraph*}(20,20)
        \fmfleft{i1}
        \fmftop{j1}
        \fmfright{o1}
        \fmf{fermion}{i1,v1,v2,o1}
        \fmf{dbl_wiggly}{v3,v2}
        \fmf{photon}{v1,v3}
        \fmf{photon}{v3,j1}
     \marrow{a}{down}{bot}{$p$}{i1,v1}
\marrow{c}{left}{lft}{$q$}{j1,v3}
\marrow{d}{down}{bot}{$p'$}{v2,o1}
\marrowdou{d}{down}{bot}{$k$}{v3,v2}
  \fmflabel{$e^-$}{i1}
        \fmflabel{$e^-$}{o1}
        \fmflabel{$\lambda$}{j1}
    \end{fmfgraph*}
 \hskip 2cm
   \begin{fmfgraph*}(20,20)
        \fmfleft{i1}
        \fmftop{j1}
        \fmfright{o1}
        \fmf{fermion}{i1,v1,v2,o1}
        \fmf{dbl_wiggly}{v1,v3}
        \fmf{photon}{v2,v3}
        \fmf{photon}{v3,j1}
     \marrow{a}{down}{bot}{$p$}{i1,v1}
\marrow{c}{left}{lft}{$q$}{j1,v3}
\marrow{d}{down}{bot}{$p'$}{v2,o1}
\marrowdou{d}{down}{bot}{$k$}{v3,v1}
  \fmflabel{$e^-$}{i1}
        \fmflabel{$e^-$}{o1}
        \fmflabel{$\lambda$}{j1}
    \end{fmfgraph*}
 \end{fmffile}
  \vspace{-3cm}
\caption{Diagrams with a two-photon vertex}
\end{figure}
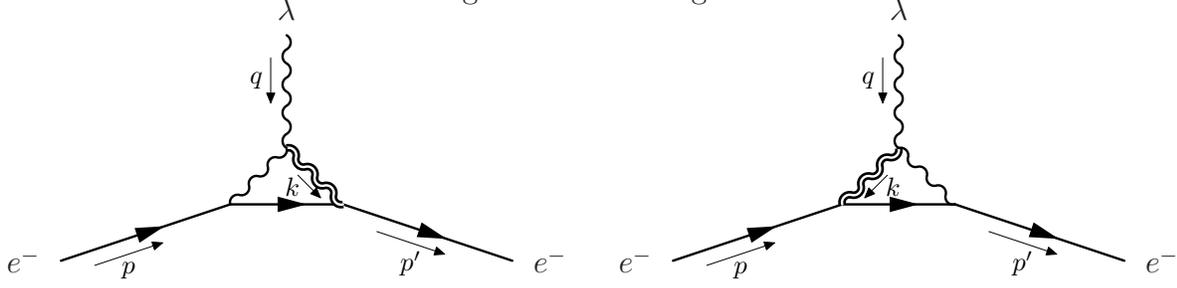

\vskip 1cm
\section{Energy-momentum tensor Ward identity and the lepton $(g-2)$-factor}

The purpose of this section is to gain some understanding of the null result obtained by  adding up the terms in (\ref{anomalous1}), (\ref{anomalous2}) and
(\ref{anomalous3}) by using the Ward identity --including contact terms, which cannot be dropped in our case because they are integrated over all spacetime-- that  expresses the conservation of the energy-momentum tensor in the quantum field theory at hand.

Let us introduce the following variations of $\bar{\psi}$, $\psi$ and $A^{\lambda}$,
\begin{equation}
\begin{array}{l}
{\delta^\nu_{x}\bar{\psi}(y)=\delta(x-y)\frac{\partial\bar{\psi}(y)}{\partial y_\nu}+\frac{i}{2}\bar{\psi(y)}S^{\nu\sigma}\frac{\partial\phantom{\bar{\psi}(y)}}{\partial y^\sigma}\delta(y-x),}\\[8pt]
{\delta^\nu_{x}\psi(y)=\delta(x-y)\frac{\partial\psi(y)}{\partial y_\nu}-\frac{i}{2}S^{\nu\sigma}\psi(y)\frac{\partial\phantom{\bar{\psi}(y)}}{\partial y^\sigma}\delta(y-x),}\\[8pt]
{\delta^\nu_{x}A^\lambda(y)=\delta(x-y)\frac{\partial A^\lambda(y)}{\partial y_\nu}-\frac{i}{2}\big(\Sigma^{\lambda}_{\mu}\big)^{\nu\sigma} A^{\mu}(y)\frac{\partial\phantom{\bar{\psi}(y)}}{\partial y^\sigma}\delta(y-x),}
\end{array}
\label{emtvariation}
\end{equation}
where $S^{\nu\sigma}$ and ${\Sigma}^{\nu\sigma}$ are the generators of the Lorentz Group in the Dirac and vector representation, respectively.

Let ${\cal O}(y)$ be a local composite operator of the fields $\bar{\psi}$, $\psi$ and $A^{\lambda}$. Let $\delta^\nu_{x}{\cal O}(y)$ its variation under the variations in (\ref{emtvariation}).

Then, by using standard path integral techniques~\cite{Suura:1974cz},  one obtains the following Ward identity in QED in Minkowski spacetime:
\begin{equation}
\begin{array}{l}
i{\langle \partial^{(x)}_{\mu} T^{\mu\nu}(x)\psi(x_2)\bar{\psi}(x_1) A^{\lambda}(x_3){\cal O}(y)\rangle=
\langle \delta^\nu_{x}\psi(x_2)\bar{\psi}(x_1) A^{\lambda}(x_3){\cal O}(y)\rangle+
\langle \psi(x_2)\delta^\nu_{x}\bar{\psi}(x_1) A^{\lambda}(x_3){\cal O}(y)\rangle}\\[8pt]
{\phantom{\langle \partial^{(x)}_{\mu} T^{\mu\nu}(x)\psi(x_2)\bar{\psi}(x_1) A^{\lambda}(x_3)}
+\langle \psi(x_2)\delta^\nu_{x}\bar{\psi}(x_1)\delta^\nu_{x} A^{\lambda}(x_3){\cal O}(y)\rangle
+\langle \psi(x_2)\delta^\nu_{x}\bar{\psi}(x_1)\delta^\nu_{x} A^{\lambda}(x_3)\delta^\nu_{x}{\cal O}(y)\rangle.}
\end{array}
\label{EMWI}
\end{equation}
This equation, and its generalizations, implements the conservation of the QED energy-momentum tensor at the quantum level. Notice that the right hand side of the Ward identity
contains the so-called contact terms, which vanish automatically only when $x$ is different from $x_1$ $x_2$, $x_3$ and $y$.

Defining
\begin{equation*}
\Delta_{\nu,\rho\sigma}(x-y)=\idkn\,e^{-ik(x-y)}\;\Big(\frac{4}{n-2}\frac{k_\nu\eta_{\rho\sigma}}{(k^2)^2}-\frac{2n}{n-2}\frac{k_\nu k_\rho k_\sigma}{(k^2)^3}\Big),
\end{equation*}
and using partial integration over $x$, it easily shown that
\begin{equation*}
\begin{array}{l}
{\frac{\kappa^2}{8}\idxonen\idxtwon\idxthreen\idxn\idyn\,e^{-i(p x_1 + q x_3-p' x_2)}}\\[8pt]
{\bar{u}(p')(i\overrightarrow{\prslash_{x_2}}-m)\langle \psi(x_2)\bar{\psi}(x_1) A^{\lambda}(x_3)
T^{\mu\nu}(x)T^{\rho\sigma}(y)\rangle(-i\overleftarrow{\prslash_{x_1}}-m)u(p)\Delta_{\mu\nu,\rho\sigma}(x-y)}
\end{array}
\end{equation*}
is equal to
\begin{equation}
\begin{array}{l}
{\frac{\kappa^2}{8}\idxonen\idxtwon\idxthreen\idxn\idyn\,e^{-i(p x_1 + q x_3-p' x_2)}}\\[8pt]
{\bar{u}(p')(i\overrightarrow{\prslash_{x_2}}-m)\langle\partial^{(x)}_{\mu} T^{\mu\nu}(x) \psi(x_2)\bar{\psi}(x_1) A^{\lambda}(x_3)
T^{\rho\sigma}(y)\rangle(-i\overleftarrow{\prslash_{x_1}}-m)u(p)\Delta_{\nu,\rho\sigma}(x-y)}
\end{array}
\label{matrixelementpart}
\end{equation}
Taking into account the Ward identity in (\ref{EMWI}), one concludes that (\ref{matrixelementpart}) is equal to
\begin{equation}
\begin{array}{l}
-i{\frac{\kappa^2}{8}\idxonen\idxtwon\idxthreen\idyn\,e^{-i(p x_1 + q x_3-p' x_2)}\;\Delta_{\nu,\rho\sigma}(x-y)}\\[8pt]
{\Big\{\bar{u}(p')(i\overrightarrow{\prslash_{x_2}}-m)\langle \psi(x_2)\bar{\psi}(x_1)\delta^\nu_{x} A^{\lambda}(x_3)T^{\rho\sigma}(y)\rangle
(-i\overleftarrow{\prslash_{x_1}}-m)u(p)}\\[8pt]
{\bar{u}(p')(i\overrightarrow{\prslash_{x_2}}-m)\langle \delta^\nu_{x}\psi(x_2)\bar{\psi}(x_1) A^{\lambda}(x_3)T^{\rho\sigma}(y)\rangle
(-i\overleftarrow{\prslash_{x_1}}-m)u(p)}\\[8pt]
{\bar{u}(p')(i\overrightarrow{\prslash_{x_2}}-m)\langle \psi(x_2)\delta^\nu_{x}\bar{\psi}(x_1) A^{\lambda}(x_3)T^{\rho\sigma}(y)\rangle
(-i\overleftarrow{\prslash_{x_1}}-m)u(p)}\\[8pt]
{\bar{u}(p')(i\overrightarrow{\prslash_{x_2}}-m)\langle \psi(x_2)\delta^\nu_{x}\bar{\psi}(x_1) A^{\lambda}(x_3)\delta^\nu_{x}T^{\rho\sigma}(y)\rangle
(-i\overleftarrow{\prslash_{x_1}}-m)u(p)\Big\}.}
\end{array}
\label{matrixelemenWI}
\end{equation}

Now, the Dirac deltas that carry the definition --see (\ref{emtvariation})-- of $\delta^{\nu}_x$ can be removed from the expression in (\ref{matrixelemenWI}) by integrating over
$x$; yielding a very lengthy expression of which we shall only render the first two summands:
\begin{equation}
\begin{array}{l}
-i{\frac{\kappa^2}{8}\idxonen\idxtwon\idxthreen\idyn\,e^{-i(p x_1 + q x_3-p' x_2)}}\\[8pt]
{\Big\{\bar{u}(p')(i\overrightarrow{\prslash_{x_2}}-m)\langle \psi(x_2)\bar{\psi}(x_1)\Delta_{\nu,\rho\sigma}(x_3-y)
\partial^{(x_3)\nu} A^{\lambda}(x_3)\,T^{\rho\sigma}(y)\rangle
(-i\overleftarrow{\prslash_{x_1}}-m)u(p)}\\[8pt]
{-\frac{i}{2}\bar{u}(p')(i\overrightarrow{\prslash_{x_2}}-m)\langle \psi(x_2)\bar{\psi}(x_1)\partial^{(x_3)}_\delta\Delta_{\nu,\rho\sigma}(x_3-y)\big(\Sigma^{\lambda}_{\mu}\big)^{\nu\delta} A^{\mu}(x_3)T^{\rho\sigma}(y)\rangle
(-i\overleftarrow{\prslash_{x_1}}-m)u(p)}\\[8pt]
{\quad\quad\quad+......\Big\}}\\[8pt]
\end{array}
\label{matrixelemenWIexpand}
\end{equation}
In the previous expression hides the one-loop difference, if any, between the Unimodular Gravity and the General Relativity contributions to the lepton $(g-2)$-factor. It is not apparent
why (\ref{matrixelemenWIexpand}), ie, (\ref{matrixelemenWI}), which comes from the contact terms in the Ward identity in (\ref{EMWI}) should yield a vanishing result for the  difference in question. The computations involved in extracting that one-loop difference from (\ref{matrixelemenWIexpand}) are tedious and involve IR divergent, at $q^\mu=0$, integrals whose final contribution is to be worked out by using the techniques displayed in the  Appendix. Since we have already shown in the previous section that the difference in question is zero indeed, we shall not take the analysis of  (\ref{matrixelemenWIexpand}) any further.

\section{Summary and discussion}

 We have shown in this paper that, either in Dimensional Regularization or in Dimensional Reduction, there is no difference between the one-loop contribution to the lepton $(g-2)$-factor that gives Unimodular Gravity and the same quantity  as given by General Relativity. This is surprising since the one-loop contribution in question, although UV finite, is not a physical observable since it depends on the regularization method. This dependence on the Regularization scheme is in total harmony with the fact that both Unimodular Gravity
and General Relativity are not renormalizable field theories and therefore any counterterm, regardless its dimension, consistent with the symmetries of the theory should be allowed.
We have shown --see (\ref{anomalous1}), (\ref{anomalous2}) and (\ref{anomalous3})-- that the contribution to the lepton $(g-2)$-factor coming for a given type of Feynman diagrams
depends on whether we are using Unimodular Gravity or General Relativity as a theory of Gravity. So the quantum equivalence for vanishing Cosmological Contant between those theories is not diagram-wise, if there exist.

We believe that the fact that in Unimodular Gravity the energy-momentum tensor only couples to the traceless part of the graviton field lies behind the zero-difference result we have obtained. This is in harmony with the derivation of Newton's Law within Unimodular  Gravity as a quantum field theory --see Ref.~\cite{Alvarez:2016uog}. Of course, if the contact --see Section 4-- terms that occur in the Ward identity expressing the conservation of the energy-momentum tensor in quantum field theory had given a net non-vanishing contribution we
would have found a non-zero-difference result.

\section{Appendix: Sample calculations}

The computation of the results in (\ref{anomalous1}), (\ref{anomalous2}) and (\ref{anomalous3}) is not straightforward, since it requires to work out, in the limit taken in (\ref{g2lfactor}), expressions made out of
Feynman integrals which develop an IR singularity at $q^\mu=0$; $q^\mu$ being the momentum of the external photon.
Actually, they are the contributions of the type in (\ref{anomalousterms}), which give rise, when carefully handled, to an IR finite result. As in plane QED the other type of contributions,ie,
\begin{equation*}
F_1(q^2)\;\bar{u}(p')\gamma^\lambda u(p),
\end{equation*}
 carry IR divergences which have to be cured by considering additional photon emissions.

 The most IR dangerous contributions  occur in the Feynman Diagrams in Figure 3, since they have an internal photon line. We shall just display the contribution in question coming from the left diagram in Figure 3. It reads
 \begin{equation}
 \begin{array}{l}
 {\idkn\,\frac{1}{((k+p)^2-m^2)(k-q)^2 (k^2)^)}\;\big\{}\\[8pt]
 {\bar{u}(p')\big[-4 m k^\lambda q\cdot k \qslash\; \kslash+2k^\lambda q^2 \kslash\;(2p\cdot k+2 m\kslash+4m^2)
 +2 m\gamma^\lambda\kslash\;(2(k\cdot q)^2-k^2 q^2)\big]u(p)\big\}.}
 \end{array}
 \label{nastyIR}
 \end{equation}
All the integrals in the previous expression are UV finite by power-counting at $n=4$, so we shall set $n=4$ for now on. Now, by introducing Feynman parameters in the standard fashion and with the help of ${\it FORM}$~\cite{Kuipers:2012rf}, one obtains that the relevant contribution coming from the expression in (\ref{nastyIR})
reads
\begin{equation}
\begin{array}{l}
{-\frac{i}{16\pi^2}\frac{1}{2}[\gamma^\lambda,\qslash]}\\[8pt]
{\int_{0}^1\,dx\;\int_{0}^{1-x}\,dy \frac{1-x-y}{(M(x,y)^2)^2}
\big\{4 m^3 q^2( - x^2+  x^2 y)+ (q^2)^2 m(- 2 y^2+ 4 x y- 4 x y^2- 3 x^2 y)\big\},
}
\end{array}
\label{xyintegrals}
\end{equation}
where
\begin{equation*}
M(x,y)^2= x^2 m^2 - q^2 y(1-x-y),\quad q^2 <0.
\end{equation*}

Because of the limit in (\ref{g2lfactor}), one needs to compute the limit $q^2\rightarrow 0^-$  of (\ref{xyintegrals}). Now, one is not allowed to set $q^2=0$ in $M(x,y)^2$ above since, then, the integral over $x$ does not exist: a signature of the generic non-smooth IR behaviour of the integrals involved in the one-loop vertex we are dealing with. To compute properly the limit $q^2\rightarrow 0^-$ of the  $(\ref{xyintegrals})$, we have shown that the follwing result hold
\begin{equation}
\begin{array}{l}
\displaystyle
{\lim_{q^2\rightarrow 0^-}\Big[q^2\int_{0}^1\,dx\;\int_{0}^{1-x}\,dy \frac{(1-x-y)x^2}{(x^2 m^2 - q^2 y(1-x-y))^2}\Big]=0,}\\[8pt]
\displaystyle
{\lim_{q^2\rightarrow 0^-}\Big[q^2\int_{0}^1\,dx\;\int_{0}^{1-x}\,dy \frac{(1-x-y)x^2 y}{(x^2 m^2 - q^2 y(1-x-y))^2}\Big]=0,}\\[8pt]
\displaystyle
{\lim_{q^2\rightarrow 0^-}\Big[(q^2)^2\int_{0}^1\,dx\;\int_{0}^{1-x}\,dy \frac{(1-x-y)y^2}{(x^2 m^2 - q^2 y(1-x-y))^2}\Big]=0,}\\[8pt]
\displaystyle
{\lim_{q^2\rightarrow 0^-}\Big[(q^2)^2\int_{0}^1\,dx\;\int_{0}^{1-x}\,dy \frac{(1-x-y)xy}{(x^2 m^2 - q^2 y(1-x-y))^2}\Big]=0,}\\[8pt]
\displaystyle
{\lim_{q^2\rightarrow 0^-}\Big[(q^2)^2\int_{0}^1\,dx\;\int_{0}^{1-x}\,dy \frac{(1-x-y)y^2}{(x^2 m^2 - q^2 y(1-x-y))^2}\Big]=0.}

\end{array}
\label{zeroinIR}
\end{equation}

To show that the results in (\ref{zeroinIR}) are correct, one may proceed as follows. Take for instance the first integral in (\ref{zeroinIR}) and perform the following change of variables when $q^2 <0$:
\begin{equation*}
x=\lambda w,\quad \lambda =\sqrt{\frac{-q^2}{m^2}}.
\end{equation*}
Then,
\begin{equation}
\begin{array}{l}
{0\leq -q^2\int_{0}^1\,dx\;\int_{0}^{1-x}\,dy \frac{(1-x-y)x^2}{(x^2 m^2 - q^2 y(1-x-y))^2}=}\\[8pt]
{\frac{\lambda}{m^2}\Big\{\int_0^1 dw\int_0^{1-\lambda w} dy \frac{(1-\lambda w -y)w^2}{(w^2+y(1-y-\lambda w))^2}+
\int_0^{1/\lambda} dw\int_0^{1-\lambda w} dy \frac{(1-\lambda w -y)w^2}{(w^2+y(1-y-\lambda w))^2}\Big\}}
\end{array}
\label{changingvariables}
\end{equation}
Now, it can be shown that
\begin{equation}
0\leq \lambda \int_0^{1/\lambda} dw\int_0^{1-\lambda w} dy \frac{(1-\lambda w -y)w^2}{(w^2+y(1-y-\lambda w))^2}\leq
 \lambda \int_0^{1/\lambda} dw w^2 f_0(w),
 \label{upbound}
 \end{equation}
 where
 \begin{equation*}
 f_0(w)=\int_0^{1} dy \frac{(1 -y)}{(w^2+y(1-y)-1/4)^2}= \frac{2w+(-1+4w^2)\text{arcotanh}(2w)}{-2w^3+8w^5}.
 \end{equation*}
Now, $0\leq w^2 f_0(\omega)$ and $ w^2 f_0(\omega)$ is (absolutely) integrable in $[1,\infty)$; so that from (\ref{upbound}), one deduces that
\begin{equation}
0\leq \lim_{\lambda\rightarrow 0^+}\Big\{\lambda \int_0^{1/\lambda} dw\int_0^{1-\lambda w} dy \frac{(1-\lambda w -y)w^2}{(w^2+y(1-y-\lambda w))^2}\Big\}\leq
\lim_{\lambda\rightarrow 0^+}\Big\{\lambda \int_0^{\infty} dw w^2 f_0(w)\Big\}=0.
 \label{zerolimitone}
 \end{equation}
Let
\begin{equation*}
\begin{array}{l}
{g_0(w,\lambda)=\int_0^{1-\lambda w} dy \frac{(1-\lambda w -y)w^2}{(w^2+y(1-y-\lambda w))^2}=}\\[8pt]
{-\frac{(1 + \lambda^2 w^2) \Big[(-1 + \lambda w) \sqrt{1 - 2 \lambda w +( 4  + \lambda^2) w^2} -2\,
       w^2 \ln\big[\frac{1 - \lambda w + \sqrt{1 - 2 \lambda w + (4  + \lambda^2) w^2}}{-1 + \lambda w +\sqrt{1 - 2 \lambda w + (4  + \lambda^2 ) w^2}}\big]\Big]}{[1 - 2 \lambda w +
     (4 + \lambda^2 w^2)]^{\frac{3}{2}}}.
}
\end{array}
\end{equation*}
It can be shown that $g_0(w,\lambda)$ is continous for all $(w,\lambda)$ in $[0,1]\times[0,\frac{1}{2^m}]$, $1/2^m\ll 1$ --recall that we are interested in values of $\lambda$ close to zero. Then, then there exists a constant $C$ such that
\begin{equation*}
0\leq g_0(w,\lambda)\leq C, \quad \forall (w,\lambda)\in [0,1]\times[0,\frac{1}{2^m}].
\end{equation*}
And, hence,
\begin{equation}
\begin{array}{l}
\displaystyle
{0\leq\lim_{\lambda\rightarrow 0^+} \Big\{\int_0^1 dw\int_0^{1-\lambda w} dy \frac{(1-\lambda w -y)w^2}{(w^2+y(1-y-\lambda w))^2}\Big\}=\lim_{\lambda\rightarrow 0^+}\Big\{\lambda \int_0^{1} dw  g_0(w,\lambda)\Big\}\leq }\\[8pt]
{\lim_{\lambda\rightarrow 0^+}\Big\{\lambda C\int_0^{1} dw\Big\}=0.}
\end{array}
\label{zerolimittwo}
\end{equation}

By taking into account (\ref{zerolimitone}) and (\ref{zerolimittwo}), one concludes that (\ref{changingvariables}) leads to
\begin{equation*}
\lim_{q^2\rightarrow 0^-}\Big\{ -q^2\int_{0}^1\,dx\;\int_{0}^{1-x}\,dy \frac{(1-x-y)x^2}{(x^2 m^2 - q^2 y(1-x-y))^2}\Big\}=0.
\end{equation*}
The remaining results in (\ref{zeroinIR}) are obtained by using the same technique.

Finally, integrals that are UV divergent are computed by using standard Dimensional Regularization techniques.

\section{Acknowledgements}
This work has been financially supported in part by the Spanish MINECO through grant
FPA2014-54154-P and MPNS COST Action MP1405. I am indebted to E. Alvarez and S. Gonzalez-Martin for valuable discussions. I thank S. Gonzalez-Martin for valuable comments on the manuscript.


\end{document}